\documentclass[aps,prd,preprint,preprintnumbers,showpacs]{revtex4-1}
\usepackage{graphicx}
\usepackage{amsmath}
\usepackage{caption}
\usepackage[section] {placeins}
\usepackage{slashed}
\usepackage[dvips]{color}
\usepackage{soul}
\usepackage{multirow}

\begin{document}

\title{Exploring the $D^*\rho$ system within QCD sum rules}
\author{A.~Mart\'inez~Torres\footnote{amartine@if.usp.br}}
 \affiliation{
Instituto de F\'isica, Universidade de S\~ao Paulo, C.P. 66318, 05389-970 S\~ao 
Paulo, SP, Brazil.
}
\author{K.~P.~Khemchandani\footnote{kanchan@if.usp.br}}
 \affiliation{
Instituto de F\'isica, Universidade de S\~ao Paulo, C.P. 66318, 05389-970 S\~ao 
Paulo, SP, Brazil.
}
\author{ M.~Nielsen\footnote{mnielsen@if.usp.br} }
 \affiliation{
Instituto de F\'isica, Universidade de S\~ao Paulo, C.P. 66318, 05389-970 S\~ao 
Paulo, SP, Brazil.
}
\author{ F.~S.~Navarra\footnote{navarra@if.usp.br}}
 \affiliation{
Instituto de F\'isica, Universidade de S\~ao Paulo, C.P. 66318, 05389-970 S\~ao 
Paulo, SP, Brazil.
}
\author{ E.~Oset\footnote{oset@ific.uv.es}}
\preprint{}

\affiliation{Departamento de F\'isica Te\'orica and IFIC, Centro Mixto Universidad de Valencia-CSIC,
Institutos de Investigaci\'on de Paterna, Apartado 22085, 46071 Valencia, Spain.}

\date{\today}

\begin{abstract}
We present a study of the $D^*\rho$ system made by using the method of QCD sum rules to determine the mass of possible resonances generated in the same system. Using isospin and spin projectors, we investigate
the different  configurations and obtain evidences for three $D^*$ mesons with isospin $I=1/2$, spin $S=0$, $1$, $2$ and with 
masses $2500\pm 67$ MeV, $2523\pm60$ MeV, and $2439\pm119$ MeV, respectively. The last state can be associated with $D^*_2(2460)$ (spin 2) listed by the Particle Data Group, while one of the first two might be related to $D^*(2640)$, with unknown spin-parity. In the case of $I=3/2$ we also find evidences of three states with spin 0, 1 and 2, respectively,
with masses $2467\pm82$ MeV, $2420\pm128$ MeV, and $2550\pm56$ MeV. The results for the sector $I=1/2$ and $S=0$, $1$, $2$, are intriguingly similar to a previous study of the  $D^*\rho$ system
based on effective field theories, supporting in this way a molecular picture for the resonances $D^*(2640)$ and $D^*_2(2460)$, while the results for $I=3/2$ hint towards the existence of exotic mesons since a multiquark configuration is required to get the quantum numbers of the states found.

\end{abstract}


\maketitle
 
\section{Introduction}
Since the discovery of the $D$ and $D^*$ mesons in the seventies by the Mark I collaboration~\cite{markI}, the interest in the charmed meson spectroscopy has grown exponentially and the newly developed
experimental facilities like Cleo, Belle, BaBar, etc., are continuously bringing more relevant information about the excited charmed meson states~\cite{cleo,belle,babar}.
This information is very useful to elucidate the way the strong interaction works when a heavy quark is involved. In this aspect, many theoretical efforts have been dedicated to the understanding
of the  properties of the excited meson states in the open as well as in the hidden charm sectors using different methods based on quark models, QCD sum rules, Lattice QCD, effective field theories, etc.,~\cite{faria,marina,moir,kol,col,daniel,laura,mdkjo,hidalgo}.

However, in spite of all the experimental and theoretical works, a rather scarce information is available on the properties of the charmed meson resonances like spin, isospin, parity, etc. 
The most clear evidence of this fact is the summary of the knowledge on these particles given by the Particle Data Group (PDG)~\cite{pdg}, which lists only six $D$ and five $D^*$ meson excited states (cataloging  separately the neutral and the charged ones) spread over the energy region of 2400-2750 MeV. Out of them, five $D$ and two $D^*$ states are omitted from the summary table due to controversy on their properties.  It is also interesting to notice that the energy region studied experimentally covers a range of only 350 MeV from the first to the last excited state, while the interaction of a $D$ or a $D^*$ meson, due to their heavy mass (around 2000 MeV), with a vector or few pseudoscalar mesons could give rise to a resonance with a mass close to 3000 MeV. Recently, a theoretical work along this line was made in Ref.~\cite{mknn}, where a system formed of a $D$ meson and the $f_0(980)$ resonance was investigated using two different techniques: one based on effective field theories and other based on QCD sum rules, and the agreement of both methods was remarkably good. As a result, a new $D$ meson state with mass around 2900 MeV was predicted. Similarly, with another recent study of the $D^*\rho$ system in s-wave using effective field theories based on hidden local symmetry~\cite{raquel} an effort has been made to shed some more light on the nature of two of the $D^*$ resonances listed by the PDG: $D^*_2(2460)$ and $D^*(2640)$. In Ref.~\cite{raquel}  the spin-parity of the former state  was confirmed to be $J^P=2^+$, and the spin and parity of the latter one was predicted to be $J^P=1^+$. In addition, a new state with mass close to 2600 MeV and spin 0, thus $J^P=0^+$, was predicted. 

In this manuscript, with the objective of understanding the properties and nature of the resonances $D^*_2(2460)$ and $D^*(2640)$, we investigate the $D^*\rho$ system using QCD sum rules to determine the mass of its possible resonances  and compare our results with the ones found in Ref.~\cite{raquel}.
\section{Framework}
We would like to investigate the generation of $D^*\rho$ molecular states using a formalism based on QCD sum rules, but before we enter into details about how to do this, let us first summarize the main aspects of the previous study of the $D^*\rho$ system made by the authors of Ref.~\cite{raquel}. This discussion will be useful while comparing the results obtained in our work with those reported in Ref.~\cite{raquel}.

\subsection{The $D^*\rho$ system in effective field theories}\label{eft}
 In Ref.~\cite{raquel}, the scattering amplitudes of the $D^*\rho$ and $D^*\omega$ coupled channel system are calculated by solving the Bethe-Salpeter equation in its on-shell factorization form~\cite{kaiser,oller,jido,hyodo}:
\begin{align}
T=(1-VG)^{-1} V.\label{BS}
\end{align}
In Eq.~(\ref{BS}),  the kernel $V$ corresponds to a $2\times2$ matrix whose elements are the lowest order amplitudes describing the transitions between the channels. These amplitudes are obtained from a Lagrangian built on the basis of the hidden local symmetry~\cite{bando}. The $2\times2$ matrix $G$ in Eq.~(\ref{BS}) is a diagonal matrix whose elements correspond to the loop function of the two hadrons present in each of the channels. As shown in Ref.~\cite{raquel}, the kernel $V$ is a function of the polarization vectors of the vector mesons, $\epsilon_\mu$, where $\mu$ is a time-spatial index. Therefore, to identify resonances arising from the dynamics involved in the interaction of the vector mesons considered, $V$ needs to be projected on the spin as well as on the isospin bases.  We can have two different isospins, $I=1/2$ or $3/2$, and three different spins, $S=0$, $1$, and $2$, for the $D^*\rho$ system. Using the phase convention $|\rho^+\rangle=-|1,1\rangle$ and $|D^{*\,0}\rangle=-|1/2,-1/2\rangle$, the isospin $1/2$ and $3/2$ states can be written in terms of the charged particles as
\begin{align}
|D^*\rho,I=1/2,I_z=1/2\rangle&=-\sqrt{\frac{2}{3}}|D^{*\,0}\rho^+\rangle+\sqrt{\frac{1}{3}}|D^{*\,+}\rho^0\rangle,\nonumber\\
|D^*\rho,I=3/2,I_z=3/2\rangle&=-|D^{*\,+}\rho^+\rangle.\label{isos}
\end{align}
Further, it was shown by the authors of Ref.~\cite{raquel} that, for small momenta of the external hadrons, the combinations
\begin{align}
P^{(0)}&=\frac{1}{3}\epsilon_i\epsilon_i\epsilon_j\epsilon_j,\nonumber\\
P^{(1)}&=\frac{1}{2}\left(\epsilon_i\epsilon_j\epsilon_i\epsilon_j-\epsilon_i\epsilon_j\epsilon_j\epsilon_i\right),\label{proj}\\
P^{(2)}&=\frac{1}{2}\left(\epsilon_i\epsilon_j\epsilon_i\epsilon_j+\epsilon_i\epsilon_j\epsilon_j\epsilon_i\right)-\frac{1}{3}\epsilon_i\epsilon_i\epsilon_j\epsilon_j,\nonumber
\end{align}
where $i$ and $j$ are spatial indices, project the kernel $V$ and, thus, the scattering matrix $T$,  on spin $0$, $1$ and $2$, respectively.

Using the amplitudes which result from the projection of the kernel $V$ on isospin and spin to solve Eq.~(\ref{BS}), it was found in Ref.~\cite{raquel} that the scattering matrix $T$ shows poles in the second Riemann sheet with masses and widths compatible with the ones of the states $D^*_2(2460)$ and $D^*(2640)$, predicting the quantum numbers of the latter one to be $J^P=1^+$. The coupling of these poles to the $D^*\rho$ channel was found to be much larger than the corresponding to the $D^*\omega$ channel, thus, leading the authors of Ref.~\cite{raquel} to conclude that the resonances $D^*_2(2460)$ and $D^*(2640)$ can be interpreted as $D^*\rho$ molecular states. In addition, the existence of a resonance with a mass around 2600 MeV and quantum numbers $J^P=0^+$ was also predicted as a consequence of the interaction of a $D^*$ with a $\rho$. 

\subsection{The $D^*\rho$ system in QCD sum rules}\label{sumrulesect}
In this manuscript, we investigate the interaction of a $D^*$ and a $\rho$ considering a different point of view than the one in Ref.~\cite{raquel}: the one of the QCD sum rules. In this case, the starting point to determine the mass of the possible  $D^*\rho$ states is the evaluation of the two-point correlation function~\cite{marina,svz,io1,rry,SNB,narison,rafael}
\begin{align}
\Pi (q^2)=i\int d^4x\, e^{iqx}\langle 0|\mathcal{T}\left[j(x)j^\dagger(0)\right]|0\rangle\label{corr},
\end{align}
where $q$ is the momentum flowing from $0$ to $\infty$, $\mathcal{T}[\cdots]$ represents the $\mathcal{T}$-ordered product and $j$ is the current associated to the $D^*\rho$ system. The key idea of the QCD sum rule method is to consider that this correlation function is of dual nature and it depends on the value of the momentum $q$. For large momentum, i.e., short distances, the correlation function represents a quark-antiquark fluctuation and can be calculated using perturbative QCD. In this case, the current $j$ is written in terms of the quark content of the $D^*$ and $\rho$ mesons. However, since we are ultimately interested in studying the properties of hadrons, the relevant energies are lower and contributions from quark condensates, gluon condensates, etc., need to be included in the evaluation of Eq.~(\ref{corr}).  This can be conveniently accomplished by using the Wilson operator product expansion (OPE) of the correlation function~\cite{wilson}. In this case,  Eq.~(\ref{corr}) is expanded in terms of all local gauge invariant operators expressible in terms of quark and gluon fields in the form of condensates and a series of coefficients. The local operators incorporate nonperturbative long-distance effects, while the coefficients, by construction, include only the short-distance domain and can be determined perturbatively. This way of evaluating the correlation function is called the ``OPE description".  In Table \ref{table} we show the values used for the different condensates and quark masses involved in the OPE description of Eq.~(\ref{corr}). Since the quark $c$ is much heavier than the quarks $u$ and $d$, we can work in the limit of massless $u$ and $d$ quarks.

\begin{table}[h!]
\caption{Values of the different parameters required for numerical 
calculations of the correlation function given by Eq.~(\ref{corr}) 
(see Refs.~\cite{marina,narison,rafael}).}\label{table}
\begin{ruledtabular}
\begin{tabular}{cc}
Parameters & Values\\
\hline
$m_c$& $1.23 \pm 0.05$ GeV\\
$\langle \bar{q} q \rangle$ & $-(0.23 \pm 0.03)^3$ GeV$^3$\\
$\langle g^2 G^2 \rangle$ & $(0.88\pm 0.25)$ GeV$^4$\\
$\langle g^3 G^3 \rangle$ & $(0.58\pm 0.18)$ GeV$^6$\\
$\langle \bar{q} \sigma \cdot G q \rangle$& 0.8$\langle \bar{q} q \rangle$ GeV$^2$\\
\end{tabular}
\end{ruledtabular}
\end{table}

At large distances, or, equivalently, small momentum, the currents $j^\dagger$ and $j$ of Eq.~(\ref{corr}) can be interpreted as operators of creation and annihilation of hadrons with the same quantum numbers as the ones of the current $j$. In this case, the correlation function is obtained by inserting a complete set of states with the same quantum numbers as those of the current under consideration. This manner of determining the correlation function is often referred to as the ``phenomenological description". The assumption made in the QCD sum rule method is that there must be a range of $q^2$ values in which both descriptions must be equivalent. Calculating the correlation function of Eq.~(\ref{corr})  using these two approaches and equating them, it is possible to obtain information about the properties of the hadronic states generated in the system.

To determine the correlation function associated to the OPE description we need to construct interpolating molecular currents for the $D^*\rho$ system. Since both $D^*$ and $\rho$ are vector mesons, the simplest current we can use for the $D^*\rho$ system is of the form
\begin{align}
j_{\mu\nu}(x)=\left[\bar {q}^1_a(x)\gamma_\mu c_a(x)\right]\left[\bar{q}^2_b(x)\gamma_\nu q^3_b(x)\right],
\end{align}
with $q_1(x)$, $q_2(x)$ and $q_3(x)$ representing the fields of the light quarks $u$ or $d$, $c(x)$ is the field for the quark $c$, $a$ and $b$ are color indices and $\gamma$ represents the Dirac matrix. Considering the isospin relations of Eq.~(\ref{isos}), the currents for the $D^*\rho$ system for the cases of total isospin $I=1/2$ and $3/2$ are thus,
\begin{align}
j^{1/2}_{\mu\nu}(x)&=-\sqrt{\frac{2}{3}}\left[(\bar{u}_a\gamma_\mu c_a)(\bar{d}_b\gamma_\nu u_b)\right.,\nonumber\\
&\left.\quad-\frac{1}{2}(\bar{d}_a\gamma_\mu c_a)(\bar{u}_b\gamma_\nu u_b-\bar{d}_b\gamma_\nu d_b)\right],\label{cur}\\
j^{3/2}_{\mu\nu}(x)&=-(\bar{d}_a\gamma_\mu c_a)(\bar{d}_b\gamma_\nu u_b),\nonumber
\end{align}
where we have omitted the $x$ dependence of the quark fields for simplicity. Using these currents, we construct the function
\begin{align}
\Pi^I_{\mu\nu,\alpha\beta}=i\int d^4x\, e^{iqx}\langle 0|\mathcal{T}\left[j^I_{\mu\nu}(x){j^I_{\alpha\beta}}^\dagger(0)\right]|0\rangle.\label{corr2}
\end{align}
The currents in Eq.~(\ref{corr2}) and, thus, $\Pi^I_{\mu\nu,\alpha\beta}$, do not have a well defined spin and need to be projected. We will elaborate on the spin projection of Eq.~(\ref{corr2}) in the next section, which will lead to a correlation function with a well defined spin, $\Pi^{I,S}(q^2)$. 

Following the standard procedure of the QCD sum rule method, contracting all the quark anti-quark pairs present in Eq.~(\ref{corr2}) through the currents in Eq.~(\ref{cur}), we can rewrite $\Pi^I_{\mu\nu,\alpha\beta}$ in terms of the quark propagators, and then we can make the OPE expansion of these propagators. In terms of the quark propagators, Eq.~(\ref{corr2}) adopts the form
\begin{align}
\Pi^I_{\mu\nu,\alpha\beta}&=i\int d^4 x\, e^{iqx}\int \frac{d^4 p}{(2\pi)^4} e^{-ipx}\nonumber\\
&\left[\textrm{Tr}\left\{\gamma_\mu S_c(p)\gamma_\alpha S_q(-x)\right\}\cdot \textrm{Tr}\left\{\gamma_\nu S_q(x)\gamma_\beta S_q(-x)\right\}\right.\nonumber\\
&\left.+A_I \textrm{Tr}\left\{\gamma_\mu S_c (p)\gamma_\alpha S_q(-x)\gamma_\nu S_q(x)\gamma_\beta S_q(-x)\right\}\right], \label{corr4}
\end{align}
where Tr$\left\{\cdots\right\}$ is the trace operator, $S_q$ represents the propagator of a light quark ($u$ or $d$), $S_c$ is the propagator of the quark $c$ and $A_I$ is a coefficient whose value is $1/2$ for isospin $1/2$ and $-1$ for $3/2$. For convenience,  we work in the momentum space for heavy quarks and in the configuration space for light quarks. The expressions for these propagators, including the terms related to quark and gluon condensates, can be found in Ref.~\cite{marina}. Making the OPE expansion for the propagators involved in Eq.~(\ref{corr4}), and projecting the result on a particular spin configuration, we can express the correlation function $\Pi^{I,S}(q^2)$ in terms of the following dispersion relation
\begin{align}
 \Pi^{I,S}_{\textrm{OPE}}(q^2)=\int_{m^2_c}^\infty ds\frac{\rho^{I,S}_{\textrm{OPE}}(s)}{s-q^2},\label{corrope}
\end{align}
with $\rho^{I,S}_{\textrm{OPE}}(s)$ being the spectral density, which is related to the imaginary part of the correlation function through $\pi \rho^{I,S}_{\textrm{OPE}}=\textrm{Im}\left[ \Pi^{I,S}_{\textrm{OPE}}\right]$. In this study of the $D^*\rho$ system we have considered condensates up to dimension 7 in the OPE description. In this way:
\begin{align}
\rho^{I,S}_{\textrm{OPE}}&=\rho^{I,S}_{\textrm{pert}}+\rho^{I,S}_{\langle\bar q q\rangle}+\rho^{I,S}_{\langle g^2 G^2\rangle}+\rho^{I,S}_{\langle\bar q g\sigma G q\rangle}+\rho^{I,S}_{{\langle\bar q q\rangle}^2}\nonumber\\
&\quad+\rho^{I,S}_{\langle g^3 G^3\rangle}+\rho^{I,S}_{\langle\bar q q\rangle\langle g^2 G^2\rangle}.\label{rho}
\end{align}
The first term in Eq.~(\ref{rho}) or perturbative term corresponds to dimension 0 in the expansion. The next term, i.e., the quark condensate contribution ($\langle\bar q q\rangle$) has dimension 3. The two gluon condensate contribution ($\langle g^2 G^2\rangle$) corresponds to dimension 4 in the expansion. The mixed condensate or $\langle\bar q g\sigma G q\rangle$ term contributes with dimension 5. The two quark and the three gluon condensates (${\langle\bar q q\rangle}^2$ and $\langle g^3 G^3\rangle$, respectively) have dimension 6 and, finally, the contribution associated to the condensate 
$\langle\bar q q\rangle\langle g^2 G^2\rangle$ has dimension 7. The result for each of the terms of Eq.~(\ref{rho}) for the different isospins and spins can be found in the Appendix of this manuscript.

In the phenomenological description,  the correlation function for the $D^*\rho$ system can be written in terms of a spectral density as
\begin{align}
 \Pi^{I,S}_{\textrm{phenom}}(q^2)=\int_{m^2_c}^\infty ds\frac{\rho^{I,S}_{\textrm{phenom}}(s)}{s-q^2}.\label{corr5}
\end{align}
All hadrons with the same quantum numbers as the ones associated to the current $j$ of Eq.~(\ref{corr}) contribute to the density of Eq.~(\ref{corr5}). Therefore, to extract information about the states we are interested in,
the spectral density $\rho^{I,S}_{\textrm{phenom}}$ needs to be parametrized conveniently. Commonly, the density of Eq.~(\ref{corr5}) is expressed as a sum of a narrow, sharp state, which represents the one we are looking for, and a smooth continuum~\cite{marina,svz,io1,rry,SNB,narison,rafael}
\begin{align}
\rho^{I,S}_{\textrm{phenom}}(s)={\lambda^2_{I,S}}\delta(s-m_{I,S}^2)+\rho^{I,S}_{\textrm{cont}}(s),
\end{align}
with $\lambda^2_{I,S}$ the coupling of the current to the state we are interested in and $m_{I,S}$ its mass. The density related to the continuum of states is assumed to vanish below a certain value of $s$, $s_0$, called continuum threshold. Above this threshold it is common to consider the ansatz~\cite{marina,svz,io1,rry,SNB,narison,rafael}
\begin{align}
\rho_{\textrm{cont}}(s)=\rho^{I,S}_{\textrm{OPE}}(s)\Theta(s-s^{I,S}_0).\label{ansatz}
\end{align}
Therefore, using this parametrization of the spectral density of Eq.~(\ref{corr5}), the correlation function in the phenomenological description adopts the form
\begin{align}
 \Pi^{I,S}_{\textrm{phenom}}(q^2)=\frac{\lambda^2_{I,S}}{m_{I,S}^2-q^2}+\int_{s^{I,S}_0}^\infty ds\,\frac{\rho^{I,S}_{\textrm{OPE}}(s)}{s-q^2}.\label{corrphen}
\end{align}
Ideally, the result from the evaluation of Eqs.~(\ref{corrope}) and (\ref{corrphen}) should be same at some range of $q^2$ at which we could just directly equate both expressions.
However, this is not completely true: the correlation function in Eq.~(\ref{corrope}) suffers from divergent contributions originated from long range interactions,
while the one in Eq.~(\ref{corrphen}) contains contributions arising from the continuum. These problems can be solved by applying the Borel transformation
to both correlation functions and then equating them, which gives the relation
\begin{align}
\lambda^2_{I,S}e^{-m_{I,S}^2/M^2}=\int_{m^2_c}^{s^{I,S}_0}ds\,\rho^{I,S}_{\textrm{OPE}}(s)e^{-s/M^2},\label{mrule}
\end{align}
where $M$ is the Borel mass parameter. Calculating the derivative of Eq.~(\ref{mrule}) with respect to $-M^{-2}$ and dividing the resulting expression by Eq.~(\ref{mrule}), we can determine the mass as
\begin{align}
m_{I,S}^2=\frac{\int_{m^2_c}^{s^{I,S}_0}ds\,s \rho^{I,S}_{\textrm{OPE}}(s)e^{-s/M^2}}{\int_{m^2_c}^{s^{I,S}_0}ds\,\rho^{I,S}_{\textrm{OPE}}(s)e^{-s/M^2}}.\label{mass}
\end{align}
Once the mass is obtained, we can calculate the coupling $\lambda_{I,S}$ through Eq.~(\ref{mrule})
\begin{align}
\lambda^2_{I,S}=\frac{\int_{m^2_c}^{s^{I,S}_0}ds\,\rho^{I,S}_{\textrm{OPE}}(s)e^{-s/M^2}}{e^{-m_{I,S}^2/M^2}}\label{lambda}
\end{align}

\section{Spin projection}
Let us discuss now how to make the spin projection of the function $\Pi^I_{\mu\nu,\alpha\beta}$ of Eq.~(\ref{corr2}). As we discussed in section~\ref{eft}, it was shown in Ref.~\cite{raquel} that the expressions given in Eq.~(\ref{proj}) project the scattering matrix on spin 0, 1, and 2 for small momentum of the external hadrons. Recalling that in quantum mechanics the Wigner-Eckart theorem establishes that all vectors (tensors) have the same matrix elements up to a reduced matrix coefficient, it is possible to show that in the non relativistic limit, taking zero momentum for the quarks, the $\vec{\gamma}$ matrices present in the currents of Eq.~(\ref{cur}) behave as the polarization vectors of Eq.~(\ref{proj}).  Therefore,
by analogy to the projectors of Eq.~(\ref{proj}), the combinations
\begin{align}
\mathcal{P}^{(0)}&=\frac{1}{3}\delta_{ij}\delta_{kl},\nonumber\\
\mathcal{P}^{(1)}&=\frac{1}{2}\left(\delta_{ik}\delta_{jl}-\delta_{il}\delta_{jk}\right),\label{delproj}\\
\mathcal{P}^{(2)}&=\frac{1}{2}\left(\delta_{ik}\delta_{jl}+\delta_{il}\delta_{jk}\right)-\frac{1}{3}\delta_{ij}\delta_{kl},\nonumber
\end{align}
with $\delta$ being the delta of Kronecker, project the currents of Eq.~(\ref{cur}) and, thus, the function $\Pi^I_{\mu\nu,\alpha\beta}$, on spin $0$, $1$, and $2$ in the limit of zero momentum for the quarks. However, this limit is not applicable to the present study. It is simple to generalize the projectors of Eq.~(\ref{delproj}) for a situation in which the quarks have a nonzero momentum. To do this, we need to establish the covariant extrapolation of the delta of Kronecker, which we call $\Delta^{\mu \nu}$. Since the delta of Kronecker is symmetric under the exchange of the two indices, $\Delta^{\mu \nu}$ should also be and to construct it we just have the metric tensor, $g^{\mu\nu}$, and the four momentum $q$. Thus, the only possibility is
 \begin{align}
 \delta_{ij}\to \Delta_{\mu\nu}\equiv-g_{\mu\nu}+\frac{q_\mu q_\nu}{q^2}.\label{Delta}
 \end{align}
As can be seen, in the limit of $\vec{q} \to \vec{0}$, $\Delta_{00}=0$, $\Delta_{i0}=0$ and $\Delta_{ij}=-g_{ij}=\delta_{ij}$. 
 
 In this form, replacing $\delta_{ij}$ for $\Delta_{\mu \nu}$ in Eq.~(\ref{delproj}) we arrive to the following projectors: 
\begin{align}
\mathcal{P}^{(0)}_\Delta&=\frac{1}{3}\Delta^{\mu\nu}\Delta^{\alpha\beta},\nonumber\\
\mathcal{P}^{(1)}_\Delta&=\frac{1}{2}\left(\Delta^{\mu\alpha}\Delta^{\nu\beta}-\Delta^{\mu\beta}\Delta^{\nu\alpha}\right),\label{DEproj}\\
\mathcal{P}^{(2)}_\Delta&=\frac{1}{2}\left(\Delta^{\mu\alpha}\Delta^{\nu\beta}+\Delta^{\mu\beta}\Delta^{\nu\alpha}\right)-\frac{1}{3}\Delta^{\mu\nu}\Delta^{\alpha\beta},\nonumber
\end{align}
which project  the currents of Eq.~(\ref{cur}) and, therefore, the function $\Pi^I_{\mu\nu,\alpha\beta}$, on spin $0$, $1$, and $2$ for any value of the four momentum $q$. Note that $\Delta^\mu_\nu\Delta^{\nu\alpha}=-\Delta^{\mu\alpha}$, which is, with the exception of the minus sign, the same normalization as the delta of Kronecker. Therefore, the normalization of the projectors of Eq.~(\ref{delproj}) also holds for the projectors of Eq.~(\ref{DEproj}). The deduction and use of spin projectors in the QCD sum rules method is, up to our knowledge, a novelty.  Although, work in this direction has been done in Ref.~\cite{hidalgo}, where operators choosing the largest spin component of a particle which couples to the corresponding current were deduced in a study related to the $X(3872)$ state.

It is easy to show, using the phenomenological description of the correlation function, that the expressions in Eq.~(\ref{DEproj}) indeed project Eq.~(\ref{corr2}) on spin 0, 1, and 2, respectively. As mentioned in section~\ref{sumrulesect} , within the phenomenological description, the calculation of the correlation function is done by inserting intermediate states for the hadron of interest. In the present case,  the current $j^I_{\mu\nu}$ of Eq.~(\ref{corr2}) couples to spin 0, 1, and 2 states, with the respective couplings defined through:
\begin{align}
\langle 0|j_{\mu\nu}|S\rangle&=\lambda_0\left(a g_{\mu\nu}+\frac{q_\mu q_\nu}{q^2}\right),\nonumber\\
\langle 0|j_{\mu\nu}|V\rangle&=\lambda^V_1\left(\epsilon_\mu\frac{q_\nu}{\sqrt{q^2}}+b\,\epsilon_\nu\frac{q_\mu}{\sqrt{q^2}}\right),\label{states}\\
\langle 0|j_{\mu\nu}|A\rangle&=\lambda^A_1\epsilon_{\mu\nu\delta\lambda}\frac{q^\delta}{\sqrt{q^2}}\epsilon^\lambda,\nonumber\\
\langle 0|j_{\mu\nu}|T\rangle&=\lambda_2\epsilon_{\mu\nu}.\nonumber
\end{align}
In Eq.~(\ref{states}) $|S\rangle$, $|V\rangle$, $|A\rangle$ and $|T\rangle$ represent a scalar, a vector, an axial and a tensor state, respectively,  $\lambda_0$, $\lambda^V_1$, $\lambda^A_1$, $\lambda_2$ are the couplings of the current to the respective states, $\epsilon^\mu$ and $\epsilon_{\mu\nu}$ are the polarization vectors  of the vector, axial and tensor states, respectively, $\epsilon_{\mu\nu\delta\lambda}$ the Levi-Civita tensor and $a$, $b$ are arbitrary constants. Saturating the function of Eq.~(\ref{corr2}) with these states and summing over the polarizations, we get the expression
 \begin{align}
 \Pi^{I}_{\mu\nu,\alpha\beta}&=\Pi^{I,0}\,\left(|a|^2g_{\mu\nu}g_{\alpha\beta}+\frac{q_\mu q_\nu q_\alpha q_\beta}{q^4}+ag_{\mu\nu}\frac{q_\alpha q_\beta}{q^2}+a^*g_{\alpha\beta}\frac{q_\mu q_\nu}{q^2}\right)\nonumber\\
 &\quad+\Pi^{I,1}_V\left(\Delta_{\mu\alpha}\frac{q_\nu q_\beta}{q^2}+b^*\Delta_{\mu\beta}\frac{q_\nu q_\alpha}{q^2}+b\Delta_{\nu\alpha}\frac{q_\mu q_\beta}{q^2}+|b|^2\Delta_{\nu\beta}\frac{q_\alpha q_\mu}{q^2}\right)\nonumber\\
 &\quad+\Pi^{I,1}_A\,\epsilon_{\mu\nu\delta\lambda}\epsilon_{\alpha\beta\gamma\sigma}\frac{q^\delta q^\gamma}{q^2}\Delta^{\lambda\sigma}+\Pi^{I,2}\,\left( \frac{1}{2}\Delta_{\mu\alpha}\Delta_{\nu\beta}+\frac{1}{2}\Delta_{\mu\beta}\Delta_{\nu\alpha}-\frac{1}{3}\Delta_{\mu\nu}\Delta_{\alpha\beta}\right),\label{expan}
 \end{align}
with $\Pi^{I,0}$, $\Pi^{I,1}_V$, $\Pi^{I,1}_A$,  and $\Pi^{I,2}$ being the parts of the correlation function associated with the scalar, vector, axial and tensor states. Now, it is easy to see that the action of $\mathcal{P}^{(0)}_\Delta$, $\mathcal{P}^{(1)}_\Delta$ and $\mathcal{P}^{(2)}_\Delta$ over Eq.~(\ref{expan}) selects the scalar, axial and tensor part, respectively, of the correlation function.

In summary, the correlation function to be used to study the $D^*\rho$ system for a certain isospin $I$ and spin $S$ is given by
 \begin{align}
 \Pi^{I,S}(q^2)=\mathcal{P}^{(S)}_\Delta\Pi^I_{\mu\nu,\alpha\beta}.\label{corr3}
 \end{align}
\section{Results}
The masses and couplings calculated with Eqs.~(\ref{mass}) and (\ref{lambda}), respectively, are functions of the Borel mass $M$. Ideally, the results should not depend on the value of the Borel mass used to determine them. But, in reality, this is not the case and one relies on the existence of a range of Borel masses (or Borel ``window") in which the results obtained for the mass and the coupling can be relied upon. The determination of this Borel window is based on the following constraints: (1) In the phenomenological description, the contribution to the correlation function arising from the pole term, which represents the state in which we are interested in, should dominate over the contribution from the continuum of states with the same quantum numbers. This condition fixes the maximum value for the Borel mass, $M_\textrm{max}$, at which the results obtained for the mass and coupling are meaningful (2). The other constraint is to guarantee the convergence of the OPE expansion. This fixes the minimum Borel mass at which the sum rule is reliable, $M_\textrm{min}$.~Here, in this manuscript, if $n$ is the maximum dimension taken into account in the OPE expansion, we consider that the right side of Eq.~(\ref{mrule}) converges when the relative contribution associated to the condensate of dimension $n-1$, i.e, the result of dividing the contribution to the integral in Eq.~(\ref{mrule}) of all terms of the series up to dimension $n-1$ by the contribution to the integral associated to all the terms,  differs, in modulus, by no more than 10-25 \% (depending on the case) from the relative contribution to the condensate of dimension n. 

Using these conditions, a valid Borel window is established, and the mass and coupling of the states for the different isospin and spin configurations can be calculated through Eqs.~(\ref{mass}) and (\ref{lambda}), respectively. Note that these results, as well as the Borel mass window, depend on the continuum threshold, $s^{I,S}_0$. This continuum threshold is a parameter of the model, but its value is not completely arbitrary. As can be seen in Eq.~(\ref{ansatz}), $\sqrt{s^{I,S}_0}$ is related to the onset of the continuum in the current $j$ under consideration and a reasonable value is about 450-500 MeV above the mass of the hadron we are looking for~\cite{marina,rafael}.  For this case, where we are searching for possible $D^*\rho$ molecular states of masses 2.450-2.600 GeV, we will use values for $\sqrt{s^{I,S}_0}\sim 3.00-3.15$ GeV.

\subsection{Isospin $I=1/2$ configuration}

\begin{figure}
\includegraphics[width=0.48\textwidth]{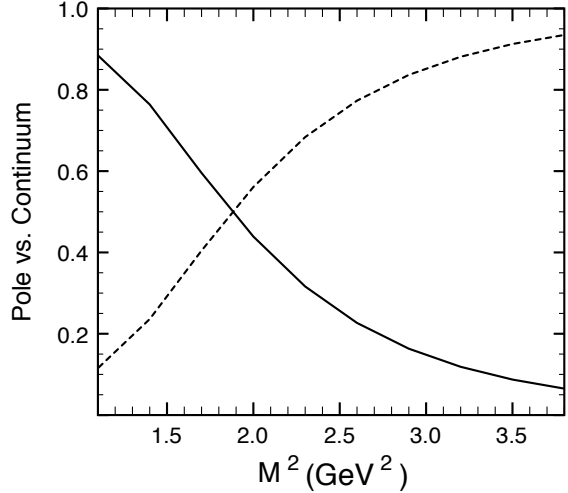}
\caption{Contributions of the pole (solid line) and the continuum (dashed line) terms (weighted by their sum) to the correlation function in the phenomenological description for the case $I=1/2$, $S=2$ as function of the squared Borel mass. These results are obtained for a value of the continuum threshold of 3.15 GeV and using the central values of Table~\ref{table} for the quark masses and condensates.}\label{polecont}
\end{figure}
\begin{figure}
\includegraphics[width=0.48\textwidth]{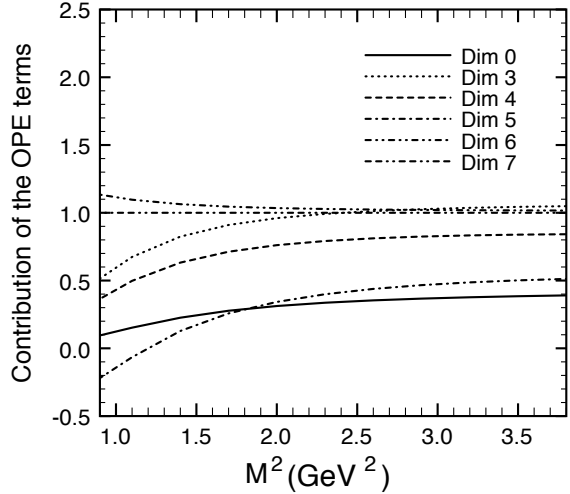}
\caption{Relative contributions of the different terms of the series for the OPE description of the correlation function for the case $I=1/2$ and $S=2$ as a function of the squared Borel mass. These results are obtained for a value of the continuum threshold of 3.15 GeV and using the central values of Table~\ref{table} for the quark masses and condensates.}\label{OPE}
\end{figure}
\begin{figure}
\includegraphics[width=0.48\textwidth]{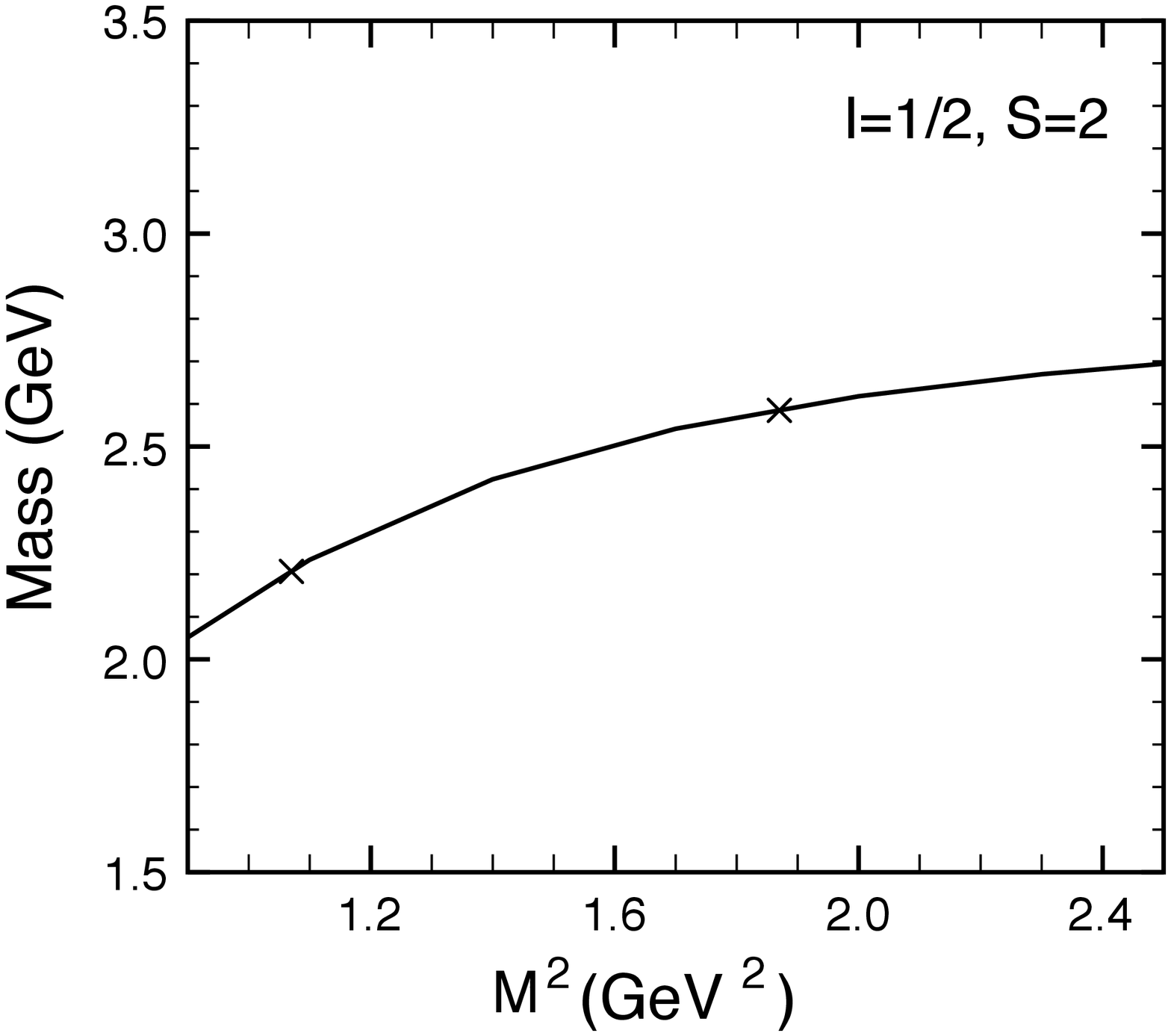}
\caption{Mass sum rule for the case $I=1/2$ and $S=2$ as a function of the squared Borel mass. The crosses in the figure indicate the Borel mass window. This result is found using a value of the continuum threshold of 3.15 GeV and considering the central values of Table~\ref{table} for the quark masses and condensates.}\label{mass2}
\end{figure}

In Fig.~\ref{polecont} we show as a function of the Borel mass and for the phenomenological description of the correlation function the contributions of the pole term and the continuum, weighted by their sum, for the case $I=1/2$, $S=2$, and a value of $\sqrt{s^{1/2,2}_0}=3.15$ GeV. As can be seen, the pole term dominates over the continuum of states for a value of the squared Borel mass of 1.87 GeV$^{2}$, thus, $M^2_\textrm{max}=1.87$ GeV$^{2}$. Similarly, for the OPE description of the correlation function, we show in Fig.~\ref{OPE} the result of the study of the convergence of the OPE series considering the central values of the quark masses and condensates of Table~\ref{table}. We find a convergence of the OPE series for a value of the squared Borel mass above 1.07 GeV$^2$, thus, $M^2_\textrm{min}=1.07$ GeV$^{2}$. Once a Borel window is found, we calculate the mass resulting from Eq.~(\ref{mass}), and obtain, as can be seen in Fig.~\ref{mass2}, that the mass sum rule is stable and gives as a result
\begin{align}
m_{1/2,2}=(2.428\pm0.151)\, \textrm{GeV}.\label{mres}
\end{align}

The value shown in Eq.~(\ref{mres}) is obtained by averaging the mass over the Borel window and by calculating the standard deviation to determine the error. Similarly, using Eq.~(\ref{lambda}), we can get the coupling of the state to the current used and we find
\begin{align}
\lambda_{1/2,2}=(8.10\pm 2.00)\times 10^{-3}\, \textrm{GeV}^5.\label{lres}
\end{align}

To estimate the uncertainty of the  results of Eqs.~(\ref{mres}) and (\ref{lres}) related to the $s^{1/2,2}_0$ value used and to the quark masses and condensates, we change $\sqrt{s^{1/2,2}_0}$ in the range 3.00-3.15 GeV and consider the range for the quark masses and condensates allowed by the error related to them (shown in Table \ref{table}). Taking into account all these sources of errors, we obtain
\begin{align}
m_{1/2,2}&=(2.439\pm0.119)\, \textrm{GeV},\nonumber\\
\lambda_{1/2,2}&=(8.14\pm 1.61 )\times 10^{-3}\, \textrm{GeV}^5.\label{m2}
\end{align}

Analogously, we repeat this process for the cases $I=1/2$ and spin $0$ and $1$, respectively. In both cases, we find the valid Borel window considering the dominance of the pole term over the continuum in the phenomenological description and the convergence of the series in the OPE description and averaging the results over the Borel window. We also change the value of the continuum threshold in the
 range $\sqrt{s^{1/2,0}_0}=\sqrt{s^{1/2,1}_0}\sim$ 3.00-3.15 GeV, and the quark masses and condensates in the range shown in Table~\ref{table}.  The results found for the masses and couplings are:
\begin{align}
m_{1/2,0}&=(2.500\pm0.067)\, \textrm{GeV},\nonumber\\
\lambda_{1/2,0}&=(3.63\pm 0.39 )\times 10^{-3}\, \textrm{GeV}^5,\nonumber\\
m_{1/2,1}&=(2.523\pm0.060)\, \textrm{GeV},\label{m01}\\
\lambda_{1/2,1}&=(6.51\pm 0.61)\times 10^{-3}\, \textrm{GeV}^5.\nonumber
\end{align}

\begin{figure}
\includegraphics[width=0.48\textwidth]{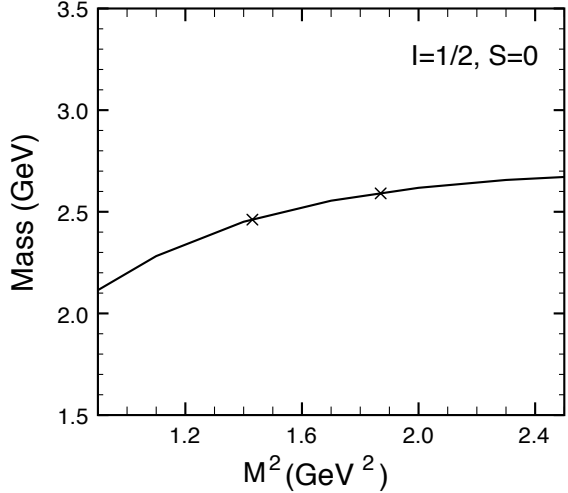}
\caption{Mass sum rule for the case $I=1/2$ and $S=0$ as a function of the squared Borel mass. The crosses in the figure indicate the Borel mass window. This result corresponds to a value of the continuum threshold of 3.15 GeV and the central values of Table~\ref{table} for the quark masses and condensates.}\label{mass0}
\end{figure}

\begin{figure}
\includegraphics[width=0.48\textwidth]{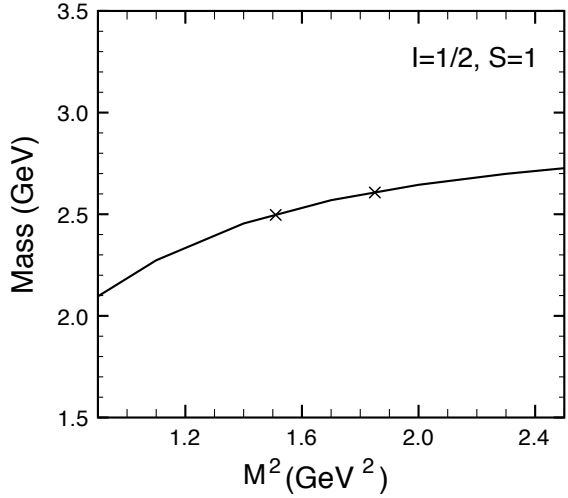}
\caption{Same as Fig.~\ref{mass0} but for $I=1/2$ and $S=1$.}\label{mass1}
\end{figure}

As an example, we show in Figs.~\ref{mass0} and \ref{mass1} the result of the mass sum rule calculation, for spin 0 and 1, respectively, for a value of $\sqrt{s^{1/2,0}_0}=\sqrt{s^{1/2,1}_0}=3.15$ GeV. The Borel mass window for both cases is indicated by crosses in the figures.

Therefore, we can conclude that three states are found for the case $I=1/2$, each with a different spin $S=0$, $1$, and $2$, with the corresponding masses and couplings given by Eqs. (\ref{m2}) and (\ref{m01}).  It is interesting to note that
the coupling of the spin 2 state to the current is bigger than the respective ones for spin 0 and 1.  These results are in striking agreement with the ones obtained in Ref.~\cite{raquel}. The state found in this manuscript with spin 2 can be associated with the $D^*_2(2450)$ listed by the PDG. For the states with spin 0 and 1, there is only one candidate listed by the PDG and that is $D^*(2640)$. However, nothing is known about the spin and parity of this state. In Ref.~\cite{raquel}, the widths related to the resonances found were calculated, obtaining  a width of around 40 MeV for the state with spin 2, 60 MeV for the state with spin 0 and practically zero width for the state with spin 1. Since the width listed by the PDG for $D^*(2640)$ is $\Gamma < 15$ MeV, the authors of Ref.~\cite{raquel} associated the state with spin 1 to $D^*(2640)$ and predicted the existence of a state with spin 0 and a similar mass, but with a much larger width. In this manuscript, we have calculated the masses and couplings for the different states. To make a proper identification of $D^*(2640)$ with one of the states with spin 0 and 1, a QCD sum rule calculation to determine the width of each of the states might be helpful. However, this is beyond the scope of the present manuscript. Thus, with the information at hand, we can only confirm the existence of two nearly degenerate states with mass around $2500\pm60$ MeV and spin 0 and 1, respectively, one of which can probably be related to $D^*(2640)$.

\subsection{Isospin $I=3/2$ configuration}
The study of the $D^*\rho$ system for the case $I=3/2$ is particularly interesting, since if a state its generated, its quantum numbers can not be obtained in terms of a quark and anti-quark configuration and we can qualify this state as an exotic one.
The study made by the authors of Ref.~\cite{raquel} disfavor the formation of molecular resonances in the $D^*\rho$ system for total isospin $3/2$. This is due to the fact that the kernel which enters in the resolution of the Bethe-Salpeter equation, Eq.~(\ref{BS}),
is repulsive for this particular isospin and for the three possible spins. Actually, this result found for the $D^*\rho$ system is similar to the one found in the study of other hadronic systems using an approach with the same spirit of Ref.~\cite{raquel}. Thus, in general, in such approaches whenever the system can generate a resonance whose quantum numbers can not be explained in terms of the constituent quark model, a repulsive interaction is found,  preventing the formation of such a state.

We have studied the configuration $I=3/2$ of the $D^*\rho$ system for the three possible spins using the QCD sum rule approach explained in Sect.~\ref{sumrulesect}. In this case, it is possible to find a valid Borel window and calculate the mass and the coupling of the corresponding states. Considering the different sources of error and averaging over the Borel window we get
\begin{align}
m_{3/2,0}&=(2.467\pm0.082)\, \textrm{GeV},\nonumber\\
\lambda_{3/2,0}&=(3.48\pm 0.48 )\times 10^{-3}\, \textrm{GeV}^5,\nonumber\\
m_{3/2,1}&=(2.420\pm 0.128)\, \textrm{GeV},\nonumber\\
\lambda_{3/2,1}&=(5.22\pm 1.12)\times 10^{-3}\, \textrm{GeV}^5,\nonumber\\
m_{3/2,2}&=(2.550\pm 0.056)\, \textrm{GeV},\nonumber\\
\lambda_{3/2,2}&=(6.60\pm 0.62)\times 10^{-3}\, \textrm{GeV}^5.\nonumber
\end{align}

The experimental search of these states will be definitely very useful to clarify the existence of possible exotic molecular hadron resonances formed due to the interaction of a $D^*$ and a $\rho$ and, at the same time, a test of the QCD sum rule approach used in this manuscript.
\section{Summary}
Using an approach based on QCD sum rules, we have studied the interaction of a $D^*$ and a $\rho$ mesons and investigated the existence of resonances for the different isospins and spins.
For the isospin $1/2$ case, we have found three states with masses $2.500\pm0.067$ GeV, $2.523\pm0.060$ GeV and $2.439\pm0.119$ GeV and spin 0, 1 and 2, respectively. The state with spin 2 can be associated with the meson $D^*_2(2450)$ listed by the PDG.  One of the other two resonances with spin $0$ or $1$ can be related to the meson $D^*(2640)$ of the PDG, whose spin-parity is unknown. For the case of isospin $3/2$, we have obtained three states with masses $2.467\pm0.082$ GeV, $2.420\pm0.128$ GeV, $2.550\pm0.056$ GeV and spin 0, 1, and 2, respectively. These states can be considered exotic in the sense that their quantum numbers can not be obtained from a quark plus antiquark configuration. While the results found for the isospin 1/2 case are in good agreement with the ones obtained by a previous study (see Ref.~\cite{raquel}) of the same system with a different approach, the results for the isospin 3/2 sector are different. The experimental studies of this system will be very relevant to confirm the results found by the model of Ref.~\cite{raquel} and the one of this manuscript in the isospin $1/2$ sector and the possible existence of $D^*\rho$ molecular resonances with isospin $3/2$ and spin 0, 1, and 2.

\section{Acknowledgements}
We thank professors Altu\u{g} \"Ozpineci and Juan Nieves for very useful discussions and for a careful reading of the manuscript. The authors would like to thank the Brazilian funding agencies FAPESP and CNPq for the financial support. This work is partly supported by the Spanish Ministerio de Economia y Competividad and European FEDER fund under the contract number FIS2011-28853-C02-01 and the Generalitat Valenciana in the program Prometeo, 2009/090. E. Oset acknowledges the support of the European Community-Research Infrastructure Integrating Activity Study of Strongly Interacting Matter (acronym HadronPhysics3, Grant Agreement n. 283286) under the Seventh Framework Programme of EU.
\appendix*
\section{OPE terms of the spectral density}\label{A}
\begin{figure*}
\centering
\includegraphics[width=0.7\textwidth]{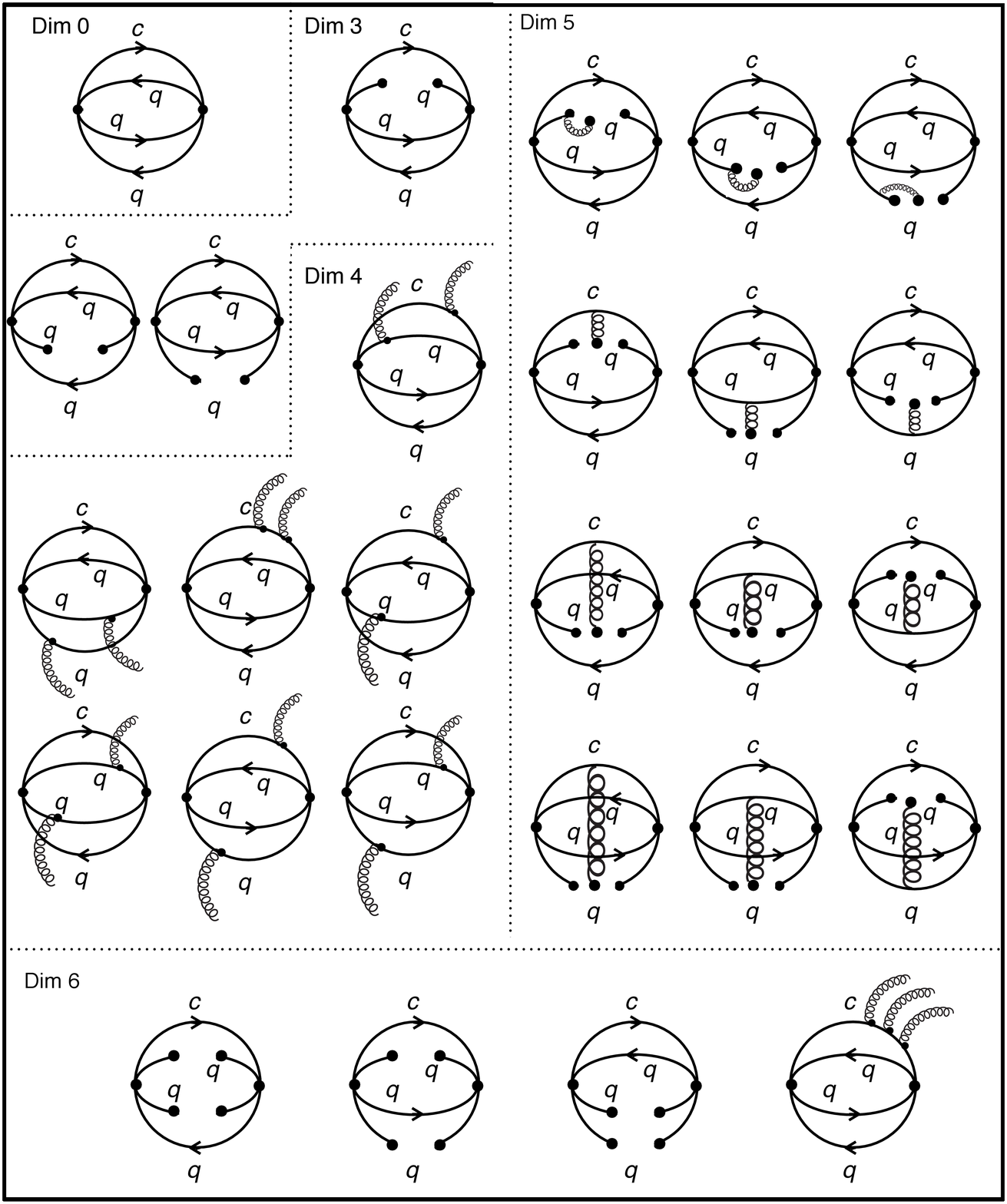}
\caption{Feynman diagrams used in the determination of the contributions up to dimension 6 for the different terms involved in the OPE series.}\label{diag1}
\end{figure*}

\begin{figure*}
\centering
\includegraphics[width=0.6\textwidth]{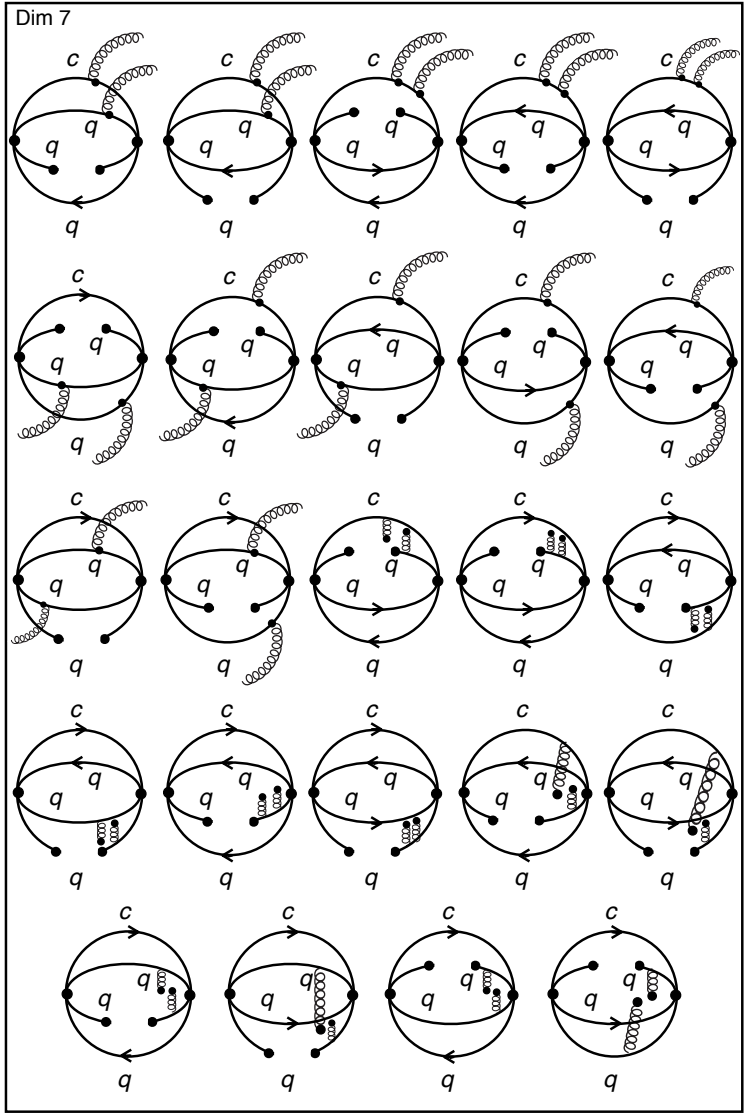}
\caption{Feynman diagrams used in the determination of the contribution associated to dimension 7 in the different terms involved in the OPE series.}\label{diag2}
\end{figure*}

In Figs.~\ref{diag1} and~\ref{diag2} we show the different Feynman diagrams considered to determine the different contributions to Eq.~(\ref{rho}).

The results for the spectral density for a certain isospin $I$, spin $S$ and dimension $D$ can be written as:
\begin{align}
\rho^{I,S}_{D}= \sum_{x=\alpha,\,\eta} \int_0^{x_0} dx\,\mathcal{R}^{I,S}_{D} (x)\,\mathcal{A}^{I,S}_{D}(x)\,\left( \sum_{k=1}^{5}\mathcal{T}^{I,S}_{D,k} (x) \mathcal{B}_k\right) \mathcal{C}_{D},\label{spec}
\end{align}
where $D=0, 3,4,5,6,7$ is a number representing the dimension of the terms considered in the OPE expansion and $\mathcal{C}_D$ is associated with the condensate of dimension $D$. In particular, the dimension $D=0$ corresponds to the perturbative term, $\mathcal{C}_0=1$. The dimension $D=3$ is related to the quark condensate or $\mathcal{C}_3=\langle\bar q q\rangle$ contribution. The two gluon condensate corresponds to $D=4$,  i.e., $\mathcal{C}_4=\langle g^2 G^2\rangle$ . The mixed condensate contributes to the dimension 5,  $\mathcal{C}_5=\langle \bar qg\sigma Gq\rangle$. For dimension $D=6$  we have contributions from the two quark condensate, ${\langle \bar q q\rangle}^2$, and the three gluon condensate,
 $\langle g^3 G^3\rangle$. The contributions to the dimension $D=7$ arise from a quark and a two gluon condensates, $\mathcal{C}_7=\langle \bar qq\rangle\langle g^2 G^2\rangle$.  The coefficients $\mathcal{B}_k$ are: $\mathcal{B}_1=m^4_c$, 
 $\mathcal{B}_2=m^2_c q^2$, $\mathcal{B}_3=q^4$, $\mathcal{B}_4=m^2_c$ and $\mathcal{B}_5=q^2$. The functions $\mathcal{R}^{I,S}_{D} (x)$ and coefficients $\mathcal{A}^{I,S}_{D}(x)$ can be found in Table~\ref{RAcoef}, while the $\mathcal{T}^{I,S}_{D,k} (x)$ functions are given in Tables~\ref{Tcoef1h} and~\ref{Tcoef3h} for isospin $I=1/2$ and $3/2$, respectively. As can be seen in these Tables, we can have two different integral variables in Eq.~(\ref{spec}): $\alpha$, in this case the integral limit $x_0=1-m^2_c/q^2$, and $\eta$, for which $x_0=1$. 
 
The results for the $\mathcal{R}^{I,S}_{D} (x)$ functions given in Table~\ref{RAcoef} are expressed in terms of the quantities defined as follows:
\begin{align}
\mathcal{R}^{I,S}_D(x)=\left\{\begin{array}{c} \mathcal{G}^{i,j,k,l}\equiv (m_c)^i \frac{\alpha^j}{(\alpha-1)^l} \left[m^2_c+(\alpha-1)q^2\right]^k,\quad\textrm{if $x=\alpha$.}\nonumber\\
\mathcal{H}^{i,j,k,l}\equiv (m_c)^i \frac{\eta^j}{(\eta-1)^l (M^2)^k} e^{\frac{m^2_c}{(\eta-1)M^2}},\quad\textrm{if $x=\eta$.}\nonumber\end{array}\right.
\end{align}
where the indices $i$, $j$, $k$, $l$ are different for each isospin, spin and dimension.

The $\mathcal{A}^{I,S}_{D}(x)$ coefficients, given in Table~\ref{RAcoef}, are written in terms of the function defined as
\begin{align}
\mathcal{F}^{i,j,k,l}&\equiv \frac{1}{5^i 3^j 2^k\pi^l}.\nonumber
\end{align}

To write the results for $\mathcal{T}^{I,S}_{D,k} (x)$ we have made used  in Tables~\ref{Tcoef1h} and~\ref{Tcoef3h} of the polynomial function 
\begin{align}
\mathcal{Y}^x_n(a_0,&\cdots,a_n)\equiv\sum_{i=0}^{n} a_i x^i.\nonumber
\end{align}
\begin{table}[ht!]
\centering
\caption{The $\mathcal{R}^{I,S}_{D}(x)$ and $\mathcal{A}^{I,S}_{D}(x)$ coefficients of Eq.~(\ref{spec}) for the different isospin and spin cases.}\label{RAcoef}
\scalebox{1}{
\begin{tabular}{|c|c|c|c|c|c|c|c|c|}
\cline{4-9}
\multicolumn{3}{c|}{}&\multicolumn{2}{|c|}{$\mathcal{R}^{I,S}_D(x)$}&\multicolumn{4}{|c|}{$\mathcal{A}^{I,S}_D(x)$}\\
\cline{4-9}
\multicolumn{3}{c|}{}&\multicolumn{2}{|c|}{$I=\frac{1}{2}$, $\frac{3}{2}$}&\multicolumn{2}{|c|}{$I=\frac{1}{2}$}&\multicolumn{2}{|c|}{$I=\frac{3}{2}$}\\
\cline{4-9}
\multicolumn{3}{c|}{}&\multicolumn{2}{|c|}{$x$}&\multicolumn{2}{|c|}{$x$}&\multicolumn{2}{|c|}{$x$}\\
\hline
$D$&OPE term&$S$&$\alpha$&$\eta$&$\alpha$&$\eta$&$\alpha$&$\eta$\\
\hline
\multirow{3}{*}{0}&\multirow{3}{*}{perturb.}&0&$\mathcal{G}^{0,3,2,3}$ &0 & $-\mathcal{F}^{1,2,16,6}$&0 & $-\mathcal{F}^{0,2,15,6}$& 0 \\
\cline{3-9}
& & 1 &$\mathcal{G}^{0,3,3,3}$ &0&$-\mathcal{F}^{0,0,13,6}$ &0 &$-\mathcal{F}^{0,0,13,6}$ &0  \\
\cline{3-9}
& & 2 &$\mathcal{G}^{0,3,2,3}$ & 0 &$-13\mathcal{F}^{1,2,15,6}$ & 0&$-\mathcal{F}^{0,2,14,6}$ &0\\
\hline
\multirow{3}{*}{3}&\multirow{3}{*}{$\langle\bar qq\rangle$}&0&$\mathcal{G}^{1,2,1,2}$ &0 &$-\mathcal{F}^{0,0,11,4}$ &0 & $-\mathcal{F}^{0,0,10,4}$&0 \\
\cline{3-9}
& & 1 &$\mathcal{G}^{1,2,1,2}$ & 0&$-\mathcal{F}^{0,1,10,4}$ & 0&$-\mathcal{F}^{0,1,9,4}$ &0 \\
\cline{3-9}
& & 2 &$\mathcal{G}^{1,2,1,2}$ & 0 &$-25\mathcal{F}^{0,1,10,4}\,$ & 0&$-5\mathcal{F}^{0,1,9,4}$ &0\\
\hline
\multirow{3}{*}{4}&\multirow{3}{*}{$\langle g^2 G^2\rangle$}&0&$\mathcal{G}^{0,1,0,3}$ &$\mathcal{H}^{2,5,0,4}$ & $-\mathcal{F}^{1,3,17,6}$&$-13\mathcal{F}^{1,3,14,6}\,$ & $-\mathcal{F}^{0,3,16,6}$&$-\mathcal{F}^{0,3,13,6}\,$  \\
\cline{3-9}
& & 1 &$\mathcal{G}^{0,1,0,3}$ &0&$-\mathcal{F}^{0,2,14,6}$ &0 &$\mathcal{F}^{0,2,13,6}$ &0 \\
\cline{3-9}
& & 2 & $\mathcal{G}^{0,1,0,3}$&$\mathcal{H}^{2,5,0,4}$ &$-\mathcal{F}^{1,3,16,6}$ & $-13\mathcal{F}^{0,3,12,6}$& $-\mathcal{F}^{0,3,15,6}$&$-\mathcal{F}^{0,3,12,6}$\\
\hline
\multirow{3}{*}{5}&\multirow{3}{*}{$\langle \bar q g\sigma G q\rangle$}&0&$\mathcal{G}^{1,1,0,1}$ &0&$\mathcal{F}^{0,2,11,4}$ & 0& $\mathcal{F}^{0,2,10,4}$&0  \\
\cline{3-9}
& & 1 &$\mathcal{G}^{1,1,0,1}$ &0&$\mathcal{F}^{0,0,11,4}$ & 0&$\mathcal{F}^{0,0,10,4}$ &0  \\
\cline{3-9}
& & 2 &$\mathcal{G}^{1,1,0,1}$ & 0& $235\mathcal{F}^{0,2,11,4}$&0 &$35\mathcal{F}^{0,2,10,4}$ &0 \\
\hline
\multirow{6}{*}{6}&\multirow{3}{*}{$\langle g^3 G^3\rangle$}&0&$\mathcal{G}^{0,3,0,3}$ &$\mathcal{H}^{4,4,1,5}$& $-\mathcal{F}^{1,2,18,6}$&$-13\mathcal{F}^{1,3,16,6}$ &$\mathcal{F}^{0,2,17,6}$ &$-\mathcal{F}^{0,3,15,6}$  \\
\cline{3-9}
& & 1 & $\mathcal{G}^{0,3,0,3}$& $\mathcal{H}^{0,4,0,4}$& $-\mathcal{F}^{0,1,15,6}$&$\mathcal{F}^{0,1,13,6}$ &$-\mathcal{F}^{0,1,15,6}$ &$\mathcal{F}^{0,1,13,6}$ \\
\cline{3-9}
& & 2 &$\mathcal{G}^{0,3,0,3}$ &$\mathcal{H}^{4,4,1,5}_{2}$&$-13\mathcal{F}^{1,2,17,6}$  &$-13\mathcal{F}^{1,3,15,6}$ &$-\mathcal{F}^{0,2,16,6}$ &$-\mathcal{F}^{0,3,14,6}$  \\
\cline{2-9}
&\multirow{3}{*}{$\langle \bar q q\rangle^2$}&0&1 &0&$-\mathcal{F}^{0,0,7,2}$ & 0&$-\mathcal{F}^{0,0,6,2}$ &0  \\
\cline{3-9}
& & 1 &1 & 0&$-\mathcal{F}^{0,1,6,2}$ &0 &$-\mathcal{F}^{0,1,5,2}$ &0 \\
\cline{3-9}
& & 2 & 1&0&$-25\mathcal{F}^{0,1,6,2}$&0 &$-5\mathcal{F}^{0,1,5,2}$ &0  \\
\hline
\multirow{3}{*}{7}&\multirow{3}{*}{$\langle \bar q q\rangle\langle g^2 G^2\rangle$}&0& $\mathcal{G}^{-1,0,0,2}$&$\mathcal{H}^{3,1,1,4}$&$\mathcal{F}^{0,3,13,4}$ & $\mathcal{F}^{0,3,13,4}$&$\mathcal{F}^{0,3,12,4}$ &
$\mathcal{F}^{0,3,12,4}$ \\
\cline{3-9}
& & 1 & $\mathcal{G}^{-1,0,0,2}$&$\mathcal{H}^{3,1,1,4}$&$-\mathcal{F}^{0,2,12,4}$ &$\mathcal{F}^{0,3,12,4}$ &$-\mathcal{F}^{0,2,11,4}$ &$\mathcal{F}^{0,3,11,4}$ \\
\cline{3-9}
& & 2 &$\mathcal{G}^{-1,0,0,2}$ &$\mathcal{H}^{3,1,1,4}$&$5\mathcal{F}^{0,3,12,4}$ & $5\mathcal{F}^{0,3,12,4}$&$-5\mathcal{F}^{0,3,11,4}\,$ &$5\mathcal{F}^{0,3,11,4}$   \\
\hline
\end{tabular}}
\end{table}


\begin{turnpage}
\begin{table}[h!]
\vspace{-2cm}
\caption{The $\mathcal{T}^{I,S}_{D,k}(x)$ coefficients of Eq.~(\ref{spec}) for the case $I=1/2$ and spin $S=0$, $1$, $2$. The absence of $x=\eta$ terms for a particular dimension indicates no $\eta$ dependence.}\label{Tcoef1h}
\scalebox{0.71}{
\begin{tabular}{|c|c|c|c|c|c|c|c|c|c|}
\cline{6-10}
\multicolumn{5}{c|}{}&\multicolumn{5}{|c|}{$\mathcal{T}^{I,S}_{D,k}(x)$}\\
\cline{6-10}
\multicolumn{5}{c|}{}&\multicolumn{5}{|c|}{$k$}\\
\hline
$I$&$S$&$D$&OPE term&$x$&1&2&3&4&5\\
\hline
\multirow{30}{*}{$\frac{1}{2}$}&\multirow{10}{*}{0}&0&perturb.&$\alpha$&$\mathcal{Y}^\alpha_2(690,-130,39)$&$2\mathcal{Y}^\alpha_3(-690,1340,-1001,351)$&
$3\mathcal{Y}^\alpha_2(1,-2,1)\mathcal{Y}^\alpha_2(230,-390,429)$ &0 &0 \\
\cline{3-10}
& &3&$\langle \bar q q\rangle$&$\alpha$&0 & 0&0& $\mathcal{Y}^\alpha_1(9,2)$&  $\mathcal{Y}^\alpha_2(-9,-1,10)$ \\
\cline{3-10}
& &\multirow{2}{*}{4}&\multirow{2}{*}{$\langle g^2 G^2\rangle$}&$\alpha$&$\mathcal{Y}^\alpha_4(2220,-6450,8590,-3915,753)\,$&$30\mathcal{Y}^\alpha_5(-148,994,-2306,2427,-1205,238)$&
$5\mathcal{Y}^\alpha_3(-1,3,-3,1)\mathcal{Y}^\alpha_3(-444,3342,-4180,1575)\,$ &0 &0 \\
\cline{5-10}
& & & &$\eta$&1 & 0&0 &0 &0 \\
\cline{3-10}
& &5&$\langle \bar q g \sigma G q\rangle$&$\alpha$&0 &0 &0 &$\mathcal{Y}^\alpha_1(79,17)$ &$\mathcal{Y}^\alpha_2(-79,28,51)$  \\
\cline{3-10}
& &\multirow{3}{*}{6}&\multirow{2}{*}{$\langle g^3 G^3\rangle$}&$\alpha$&0 & 0& 0& $\mathcal{Y}^\alpha_2(690,-130,39)$ &  $5\mathcal{Y}^\alpha_3(-46,72,-39,13)\,$ \\
\cline{5-10}
& & & &$\eta$&0 &0 &0 &$\mathcal{Y}^\eta_1(0,2)$ $-$ $\mathcal{Y}^\eta_2(10,-11,1)\frac{M^2}{m^2_c}$ & 0\\
\cline{4-10}
& &&${\langle \bar q q\rangle}^2$&$\alpha$&0 &0 &0 &$\mathcal{Y}^\alpha_1(3,2)$ &$3\mathcal{Y}^\alpha_2(-1,-1,2)$\\
\cline{3-10}
&  &\multirow{2}{*}{7} & \multirow{2}{*}{$\langle \bar q q\rangle\langle g^2 G^2\rangle$}&$\alpha$&0 &0 & 0&$\mathcal{Y}^\alpha_3(185,-384,108,-8)$ & 0\\
\cline{5-10}
& & & &$\eta$& 0&0 &0 & $\mathcal{Y}^\eta_2(0,0,12)$ $+$ $\mathcal{Y}^\eta_3(-20,47,-49,22)\frac{M^2}{m^2_c}$&0\\
\cline{2-10}
&\multirow{10}{*}{1}&0&perturb.&$\alpha$&0&0&0&$\mathcal{Y}^\alpha_1(-2,1)$ &$\mathcal{Y}^\alpha_2(2,-11,9)$ \\
\cline{3-10}
& &3&$\langle \bar q q\rangle$&$\alpha$&0 & 0&0& $\mathcal{Y}^\alpha_1(42,11)$&  $\mathcal{Y}^\alpha_2(-42,-13,55)$ \\
\cline{3-10}
& &4&$\langle g^2 G^2\rangle$&$\alpha$&$\mathcal{Y}^\alpha_3(-114,183,-89,14)$&$6\mathcal{Y}^\alpha_4(38,-103,102,-48,11)$ &$\mathcal{Y}^\alpha_3(-1,3,-3,1)\mathcal{Y}^\alpha_2(114,-93,40)$ &0 &0 \\
\cline{3-10}
& &5&$\langle \bar q g \sigma G q\rangle$&$\alpha$&0 &0 &0 &$\mathcal{Y}^\alpha_1(28,9)$ &$\mathcal{Y}^\alpha_2(-28,1,27)$  \\
\cline{3-10}
& &\multirow{3}{*}{6}&\multirow{2}{*}{$\langle g^3 G^3\rangle$}&$\alpha$&0 & 0& 0& $3\mathcal{Y}^\alpha_1(-2,1)$ &  $\mathcal{Y}^\alpha_2(2,-5,3)$ \\
\cline{5-10}
& & & &$\eta$&$\mathcal{Y}^\eta_4(0,0,0,0,1)$ &0 &0 &0 & 0\\
\cline{4-10}
& &&${\langle \bar q q\rangle}^2$&$\alpha$&0 &0 &0 &$\mathcal{Y}^\alpha_1(10,13)$ &$\mathcal{Y}^\alpha_2(-10,-29,39)$\\
\cline{3-10}
&  &\multirow{2}{*}{7} & \multirow{2}{*}{$\langle \bar q q\rangle\langle g^2 G^2\rangle$}&$\alpha$&0 &0 & 0&$\mathcal{Y}^\alpha_3(-98,231,-120,40)$ & 0\\
\cline{5-10}
& & & &$\eta$& 0&0 &0 & $\mathcal{Y}^\eta_2(0,0,22)$ $+$ $\mathcal{Y}^\eta_3(-114,444,-581,251)\frac{M^2}{m^2_c}$&0\\
\cline{2-10}
&\multirow{10}{*}{2}&0&perturb.&$\alpha$&$\mathcal{Y}^\alpha_2(30,5,3)$&$2\mathcal{Y}^\alpha_3(-30,5,-2,27)$&
$3\mathcal{Y}^\alpha_2(1,-2,1)\mathcal{Y}^\alpha_2(10,15,33)$ &0 &0 \\
\cline{3-10}
& &3&$\langle \bar q q\rangle$&$\alpha$&0 & 0&0& $\mathcal{Y}^\alpha_1(3,1)$&  $\mathcal{Y}^\alpha_2(-3,-2,5)$ \\
\cline{3-10}
& &\multirow{2}{*}{4}&\multirow{2}{*}{$\langle g^2 G^2\rangle$}&$\alpha$&$\mathcal{Y}^\alpha_4(-5700,5250,4330,-3645,753)$&$60\mathcal{Y}^\alpha_5(190,-145,-663,1094,-595,119)$ &$25\mathcal{Y}^\alpha_3(-1,3,-3,1)\mathcal{Y}^\alpha_3(228,546,-860,315)$ &0 &0 \\
\cline{5-10}
& & & &$\eta$&1 &0 &0 &0 &0 \\
\cline{3-10}
& &5&$\langle \bar q g \sigma G q\rangle$&$\alpha$&0 &0 &0 &$\mathcal{Y}^\alpha_1(2,1)$ &$\mathcal{Y}^\alpha_2(-2,-1,3)$  \\
\cline{3-10}
& &\multirow{3}{*}{6}&\multirow{2}{*}{$\langle g^3 G^3\rangle$}&$\alpha$&0 & 0& 0& $\mathcal{Y}^\alpha_2(30,5,3)$ &  $5\mathcal{Y}^\alpha_3(-2,1,0,1)$ \\
\cline{5-10}
& & & &$\eta$&0 &0 &0 &$\mathcal{Y}^\eta_1(0,2)$ $-$ $\mathcal{Y}^\eta_2(-5,4,1)\frac{M^2}{m^2_c}$ & 0\\
\cline{4-10}
& &&${\langle \bar q q\rangle}^2$&$\alpha$&0 &0 &0 &$\mathcal{Y}^\alpha_1(1,1)$ &$\mathcal{Y}^\alpha_2(-1,-2,3)$\\
\cline{3-10}
&  &\multirow{2}{*}{7} & \multirow{2}{*}{$\langle \bar q q\rangle\langle g^2 G^2\rangle$}&$\alpha$&0 &0 & 0&$\mathcal{Y}^\alpha_3(109,-261,156,-64)$ & 0\\
\cline{5-10}
& & & &$\eta$& 0&0 &0&$\mathcal{Y}^\eta_2(0,0,10)$ $+$ $\mathcal{Y}^\eta_3(-94,301,-340,133)\frac{M^2}{m^2_c}$ & 0\\
\hline
\end{tabular}}
\end{table}
\end{turnpage}

\begin{turnpage}
\begin{table}[h!]
\vspace{-2cm}
\caption{The $\mathcal{T}^{I,S}_{D,k}(x)$ coefficients of Eq.~(\ref{spec}) for the case $I=3/2$ and spin $S=0$, $1$, $2$. The absence of $x=\eta$ terms for a particular dimension indicates no $\eta$ dependence.}\label{Tcoef3h}
\scalebox{0.71}{
\begin{tabular}{|c|c|c|c|c|c|c|c|c|c|}
\cline{6-10}
\multicolumn{5}{c|}{}&\multicolumn{5}{|c|}{$\mathcal{T}^{I,S}_{D,k}(x)$}\\
\cline{6-10}
\multicolumn{5}{c|}{}&\multicolumn{5}{|c|}{$k$}\\
\hline
$I$&$S$&$D$&OPE term&$x$&1&2&3&4&5\\
\hline
\multirow{30}{*}{$\frac{3}{2}$}&\multirow{10}{*}{0}&0&perturb.&$\alpha$&$\mathcal{Y}^\alpha_2(78,-10,3)$&$2\mathcal{Y}^\alpha_3(-78,128,-77,27)$&
$3\mathcal{Y}^\alpha_2(1,-2,1)\mathcal{Y}^\alpha_2(26,-30,33)$ &0 &0 \\
\cline{3-10}
& &3&$\langle \bar q q\rangle$&$\alpha$&0 & 0&0& $\mathcal{Y}^\alpha_1(3,2)$&  $\mathcal{Y}^\alpha_2(-3,-7,10)$ \\
\cline{3-10}
& &\multirow{2}{*}{4}&\multirow{2}{*}{$\langle g^2 G^2\rangle$}&$\alpha$&$\mathcal{Y}^\alpha_4(-12,-222,586,-243,33)\,$&$6\mathcal{Y}^\alpha_5(4,230,-718,759,-325,50)$&
$\mathcal{Y}^\alpha_3(-1,3,-3,1)\mathcal{Y}^\alpha_3(12,1194,-1220,315)\,$ &0 &0 \\
\cline{5-10}
& & & &$\eta$&1 & 0&0 &0 &0 \\
\cline{3-10}
& &5&$\langle \bar q g \sigma G q\rangle$&$\alpha$&0 &0 &0 &$\mathcal{Y}^\alpha_1(29,37)$ &$\mathcal{Y}^\alpha_2(-29,-82,111)$  \\
\cline{3-10}
& &\multirow{3}{*}{6}&\multirow{2}{*}{$\langle g^3 G^3\rangle$}&$\alpha$&0 & 0& 0& $\mathcal{Y}^\alpha_2(-78,10,-3)$ &  $\mathcal{Y}^\alpha_3(26,-36,15,-5)\,$ \\
\cline{5-10}
& & & &$\eta$&0 &0 &0 &$\mathcal{Y}^\eta_1(0,2)$ $-$ $\mathcal{Y}^\eta_2(10,-11,1)\frac{M^2}{m^2_c}$ & 0\\
\cline{4-10}
& &&${\langle \bar q q\rangle}^2$&$\alpha$&0 &0 &0 &$\mathcal{Y}^\alpha_1(1,2)$ &$\mathcal{Y}^\alpha_2(-1,-5,6)$\\
\cline{3-10}
&  &\multirow{2}{*}{7} & \multirow{2}{*}{$\langle \bar q q\rangle\langle g^2 G^2\rangle$}&$\alpha$&0 &0 & 0&$\mathcal{Y}^\alpha_3(139,-372,324,-136)$ & 0\\
\cline{5-10}
& & & &$\eta$& 0&0 &0 & $\mathcal{Y}^\eta_2(0,0,12)$ $+$ $\mathcal{Y}^\eta_3(-196,637,-719,278)\frac{M^2}{m^2_c}$&0\\
\cline{2-10}
&\multirow{10}{*}{1}&0&perturb.&$\alpha$&0&0&0&$\mathcal{Y}^\alpha_1(-2,1)$ &$\mathcal{Y}^\alpha_2(2,-11,9)$ \\
\cline{3-10}
& &3&$\langle \bar q q\rangle$&$\alpha$&0 & 0&0& $\mathcal{Y}^\alpha_1(12,7)$&  $\mathcal{Y}^\alpha_2(-12,-23,35)$ \\
\cline{3-10}
& &4&$\langle g^2 G^2\rangle$&$\alpha$&$\mathcal{Y}^\alpha_3(48,-60,10,5)$&$3\mathcal{Y}^\alpha_4(-32,64,-18,-27,13)$ &$4\mathcal{Y}^\alpha_3(-1,3,-3,1)\mathcal{Y}^\alpha_2(-12,-3,10)$ &0 &0 \\
\cline{3-10}
& &5&$\langle \bar q g \sigma G q\rangle$&$\alpha$&0 &0 &0 &$\mathcal{Y}^\alpha_1(8,9)$ &$\mathcal{Y}^\alpha_2(-8,-19,27)$  \\
\cline{3-10}
& &\multirow{3}{*}{6}&\multirow{2}{*}{$\langle g^3 G^3\rangle$}&$\alpha$&0 & 0& 0& $3\mathcal{Y}^\alpha_1(-2,1)$ &  $\mathcal{Y}^\alpha_2(2,-5,3)$ \\
\cline{5-10}
& & & &$\eta$&1 &0 &0 &0 & 0\\
\cline{4-10}
& &&${\langle \bar q q\rangle}^2$&$\alpha$&0 &0 &0 &$\mathcal{Y}^\alpha_1(8,5)$ &$\mathcal{Y}^\alpha_2(-8,-7,15)$\\
\cline{3-10}
&  &\multirow{2}{*}{7} & \multirow{2}{*}{$\langle \bar q q\rangle\langle g^2 G^2\rangle$}&$\alpha$&0 &0 & 0&$\mathcal{Y}^\alpha_3(-64,147,-96,32)$ & 0\\
\cline{5-10}
& & & &$\eta$& 0&0 &0 & $\mathcal{Y}^\eta_2(0,0,14)$ $+$ $\mathcal{Y}^\eta_3(-210,582,-571,199)\frac{M^2}{m^2_c}$&0\\
\cline{2-10}
&\multirow{10}{*}{2}&0&perturb.&$\alpha$&$\mathcal{Y}^\alpha_2(30,5,3)$&$2\mathcal{Y}^\alpha_3(-30,5,-2,27)$&
$3\mathcal{Y}^\alpha_2(1,-2,1)\mathcal{Y}^\alpha_2(10,15,33)$ &0 &0 \\
\cline{3-10}
& &3&$\langle \bar q q\rangle$&$\alpha$&0 & 0&0& $\mathcal{Y}^\alpha_1(3,1)$&  $\mathcal{Y}^\alpha_2(-3,-2,5)$ \\
\cline{3-10}
& &\multirow{2}{*}{4}&\multirow{2}{*}{$\langle g^2 G^2\rangle$}&$\alpha$&$\mathcal{Y}^\alpha_4(-372,354,250,-189,33)$&$12\mathcal{Y}^\alpha_5(62,-53,-183,292,-143,25)$ &$\mathcal{Y}^\alpha_3(-1,3,-3,1)\mathcal{Y}^\alpha_3(372,834,-1100,315)$ &0 &0 \\
\cline{5-10}
& & & &$\eta$&1 &0 &0 &0 &0 \\
\cline{3-10}
& &5&$\langle \bar q g \sigma G q\rangle$&$\alpha$&0 &0 &0 &$\mathcal{Y}^\alpha_1(2,1)$ &$\mathcal{Y}^\alpha_2(-2,-1,3)$  \\
\cline{3-10}
& &\multirow{2}{*}{6}&\multirow{2}{*}{$\langle g^3 G^3\rangle$}&$\alpha$&0 & 0& 0& $\mathcal{Y}^\alpha_2(30,5,3)$ &  $5\mathcal{Y}^\alpha_3(-2,1,0,1)$ \\
\cline{5-10}
& & & &$\eta$&0 &0 &0 &$\mathcal{Y}^\eta_1(0,2)$ $-$ $\mathcal{Y}^\eta_2(-5,4,1)\frac{M^2}{m^2_c}$ & 0\\
\cline{4-10}
& &&${\langle \bar q q\rangle}^2$&$\alpha$&0 &0 &0 &$\mathcal{Y}^\alpha_1(1,1)$ &$\mathcal{Y}^\alpha_2(-1,-2,3)$\\
\cline{3-10}
&  &\multirow{2}{*}{7} & \multirow{2}{*}{$\langle \bar q q\rangle\langle g^2 G^2\rangle$}&$\alpha$&0 &0 & 0&$\mathcal{Y}^\alpha_3(-53,117,-60,8)$ & 0\\
\cline{5-10}
& & & &$\eta$& 0&0 &0&$\mathcal{Y}^\eta_2(0,0,2)$ $+$ $\mathcal{Y}^\eta_3(-14,41,-44,17)\frac{M^2}{m^2_c}$ & 0\\
\hline
\end{tabular}}
\end{table}
\end{turnpage}
\FloatBarrier

\end{document}